\documentclass[10pt,a4paper]{IEEEtran}
\usepackage{times}
\usepackage{cmap}
\usepackage{cite}
\usepackage{amsmath,amssymb,subcaption,amsthm,graphicx}
\usepackage[utf8]{inputenc}
\usepackage[font=footnotesize]{caption}
\usepackage{booktabs}
\usepackage[utf8]{inputenc}
\usepackage[T1]{fontenc}
\usepackage{color}
\usepackage[mathcal]{euscript}
\usepackage{cuted}

\usepackage{multirow}
\usepackage{siunitx}

\usepackage{stfloats}
\begin{document}

\title{OFDM with Index Modulation in Orbital Angular Momentum Multiplexed Free Space Optical Links   }

\author{\IEEEauthorblockA{El-Mehdi Amhoud\IEEEauthorrefmark{1},
Marwa Chafii\IEEEauthorrefmark{2},
Ahmad Nimr\IEEEauthorrefmark{3},
Gerhard Fettweis\IEEEauthorrefmark{3}
}

\IEEEauthorblockA{\IEEEauthorrefmark{1} School of Computer and Communication Sciences,  Mohammed VI Polytechnic University, Morocco \\
\IEEEauthorrefmark{2} ETIS, UMR8051,
CY Cergy Paris Université, ENSEA, CNRS, France\\
\IEEEauthorrefmark{3}Vodafone Chair Mobile Communication Systems, Technische Universitat Dresden, Germany \\
Emails: elmehdi.amhoud@um6p.ma, 
marwa.chafii@ensea.fr, \{ahmad.nimr,
gerhard.fettweis\}@ifn.et.tu-dresden.de}}

\maketitle

\begin{abstract}
Communication using orbital angular momentum (OAM) modes has recently received a considerable interest in free space optical (FSO) communications. Propagating OAM modes through free space may be subject to atmospheric turbulence (AT) distortions that cause signal attenuation and crosstalk which degrades the system capacity and increases the error probability.
In this paper, we propose to enhance the OAM FSO communications in terms of bit error rate and spectral efficiency, for different levels of AT regimes. The performance gain is achieved by introducing orthogonal frequency division multiplexing (OFDM) with index modulation technique to the OAM FSO system.

\end{abstract}

\IEEEpeerreviewmaketitle

\section{Introduction}
  
Orbital angular momentum (OAM) has been proposed  to transmit multiple signals  over  free space optical (FSO) channels \cite{zhao2015capacity,zhang2016capacity}. This simultaneous transmission of information on OAM modes is possible thanks to the orthogonality property of OAM beams that allows propagation without interference between signals. A simple transmission scheme of OAM-FSO systems consists in using intensity-modulation at the transmitter along with a direct-detection at the receiver \cite{amhoudTWC2019}. This configuration has a low cost and can be more feasible for practical deployment. On the other hand, coherent detection mostly used  for optical fibers can also be used for OAM-FSO systems. In coherent systems, both the intensity and phase of the received signal are detected, thus the spectral efficiency (SE) can be further increased by using higher order modulation formats \cite{yang2017turbulence}. However, these benefits come at the cost of high expense of the receiver and a complexity that increases with the number of modes. 
By using OAM multiplexing as an additional degree of freedom to polarization and wavelength, coherent transmissions were demonstrated to  achieve more than 1 Pbit/s capacity using 26 modes  in laboratory experiments \cite{1Pbits}.
	
In real life communication scenarios, transmitted OAM beams are subject to several impairments that if not properly addressed and compensated may significantly degrades the  FSO transmission performance. These impairments include misalignment between the transmitter and the receiver and also atmospheric turbulence (AT). We focus on this work in OAM transmission systems affected by AT distortions, in which the refractive index of the air experiences spatial variations.
The propagation of  OAM beams in turbulent atmosphere leads to wavefront distortions as well as beam spread and wandering. Moreover, the power of a signal carried by a particular OAM mode  spreads to other modes   resulting in modal crosstalk. 

Different compensation techniques have been proposed to mitigate AT. Compensation can be done  at the beam level by correcting the optical wavefront, this technique is known as adaptive optics (AO). The principle of AO consists of sensing the deformations of an incoming wavefront and then applying appropriate corrections. In practice, this can be realized by deforming a mirror, or by controlling a liquid crystal array. A reduction in crosstalk by more than 12 dB was shown in \cite{Ren} by applying both pre- and post-compensation based on AO. 
An alternative way to mitigate the AT effect consists in applying digital signal processing (DSP) techniques such as channel coding, coded modulation, or equalization. In \cite{amhoud2018oam}, space-time coding  with appropriate OAM mode selection  was demonstrated to absorb all  penalties for weak AT and considerable codings gains were shown in the strong AT regime. In \cite{CodingFSOMIMO}, MIMO equalization associated with heterodyne detection was shown to mitigate turbulence-induced crosstalk for 4 OAM beams carrying 20 Gbit/s QPSK signals. In \cite{sunOFDM}, OFDM along with a pilot assisted least square algorithm was shown to reduce the modal crosstalk which improves the bit-error rate (BER) by two orders of magnitude. In fact, OFDM can efficiently reduce inter-symbol interference due to its robustness against frequency selective fading which can also mitigate the random fading effect caused by atmospheric turbulence in the FSO channel.

In this work, we propose orthogonal frequency division multiplexing with index modulation (OFDM-IM) to improve the spectral efficiency  of OAM-FSO transmission systems in the presence of atmospheric turbulence. Moreover, we also show that OFDM-IM brings significant coding gains compared to the classical OFDM for both the weak and strong atmospheric turbulence regimes.   
\par 
This paper is organized as follows: In Section II, we start by describing the propagation of OAM modes in atmospheric turbulence. In Section III, we define the OFDM-IM OAM-FSO system and show the spectral efficiency  increase. In Section IV, we compare the BER performance of the OFDM and OFDM-IM schemes. Section V, concludes and sets forth the perspectives of our work.
\begin{figure*}[t]
	\centering
	\includegraphics[width=16cm,height=7cm]{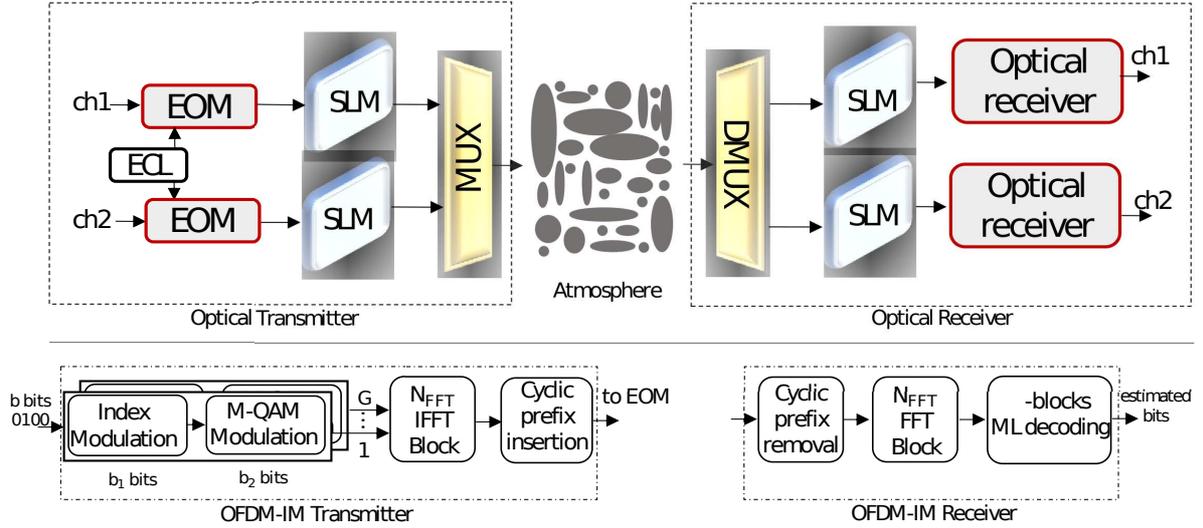}
	\caption{Top: Optical transmitter/receiver front ends: ECL: External cavity laser, EOM: Electro-optical modulator, SLM: Spatial light modulator, MUX/DEMUX: Multiplexer/Demultiplexer. Bottom: OFDM-IM transmitter/receiver architecture.}
	\label{channel}
\end{figure*}
\section{Propagation of OAM Modes }
\subsection{Orbital Angular Momentum Multiplexing}
A lightwave carrying an OAM mode of topological charge $m$ is a wave having a helical phase-front induced by an azimuthally varying phase term $\exp(im\phi)$, where $\phi$ is the azimuth. To realize OAM multiplexing, single and superpositions of orthogonal beams that have a well-defined vorticity such as the  Hermite-Gaussian (HG) or the  Laguerre-Gauss (LG) beams can be used. Here, we consider OAM carrying beams derived from LG modes. The field distribution of the LG beams is given by \cite{Doster}:
\begin{multline}
u\left ( r,\phi ,z \right )= \sqrt{\frac{2p!}{\pi(p+\left | m \right |)!}}\frac{1}{w(z)}\left [ \frac{r\sqrt{2}}{w(z)} \right ]^{\left | m \right |}\\
\times L^m_p  \left [ \frac{2r^2}{w^2(z)} \right ]\times 
\exp\left (\frac{-r^2}{w^2(z)}  \right ) \times 
\exp\left ( \frac{-ikr^2z}{2(z^2+z^2_R)} \right ) \\ 
\times \exp\left ( i(2p + \left | m \right | + 1)) 
\tan^{-1}\left ( z/z_R \right ) \right ) \times \exp\left ( -im\phi \right ),
\end{multline}
where $r$ refers to  the radial distance, $\phi$ is the azimuth angle, $z$ is the propagation distance. \mbox{ $w\left ( z \right )=w_0\sqrt{\left ( 1+\left ( z/z_R \right )^2 \right )}$} is the beam radius at the distance $z$, where $w_0$ is the beam waist of   the Gaussian beam, $z_R=\pi w^2_0/\lambda$ is the Rayleigh range, and $\lambda$ is the optical carrier \cite{Doster}.  $k=2\pi/ \lambda$ is the propagation constant. $ L^m_p(\cdot )$ is the generalized Laguerre polynomial, where $p$ and $m$ represent the radial and angular mode numbers. OAM modes correspond to the subset of LG modes having $p=0$ and $m\neq 0$.
\newline\indent
Generation of OAM beams in practice can be realized through different techniques  including spiral phase plates (SPP), q-plates, metamaterials, and computer-generated holograms (CGHs) loaded on spatial light modulators (SLMs) \cite{heckenberg1992generation}. An SPP is an optical element having the form of spiral staircases that shapes an incident Gaussian beam into a twisted beam having a helical phase-front. A single SPP allows the generation of a unique OAM mode with a particular topological charge in a stable and efficient manner for a particular wavelength which is similar to q-plates that is usually made with liquid crystals with strong wavelength dependence. Metamaterials-based devices and integrated devices are still limited to a low number of OAM modes and have not yet been used in communication scenarios. However, an SLM can be dynamically addressed to change a digital hologram displayed on a liquid crystal display (LCD) to generate single and superposition of OAM beams in a wide wavelength range from a Gaussian incident beam. SLMs are the most commonly used devices in communication experiments involving OAM beams in different wavelength ranges. At the receiver side, the inverse operation can be performed using the same device to transform an incoming OAM mode back to a Gaussian beam. The idea is to apply an optical scalar product measurement between the incident OAM beam and a CGH with the conjugate phase at the image plane of a Fourier transforming lens. The only inconvenience of using such devices is the diffraction losses at the transmission and reception due to the efficiency of the LCD. The produced Gaussian beam can then be injected into a photodetector to recover the originally encoded signal.
The vorticity of OAM beams propagating in  a FSO media without  atmospheric turbulence is preserved and OAM beams maintain orthogonality as they propagate which can be described by:
\begin{equation}
\int  u_m(r,\phi,z) u_n^\ast (r,\phi,z)rdr d\phi=\left\{\begin{matrix}
1~,~~\text{if}~ m=n\\ 
0~,~~\text{if}~ m\neq n
\end{matrix}\right.,
\end{equation}
where $u_m(\boldsymbol{r},z)$ refers to the normalized field distribution of OAM mode of order $m$ at distance $z$ and $\boldsymbol{r}$ refers to the radial position vector.  
\subsection{ OAM Propagation in Turbulence}
The propagation of OAM modes can be affected by atmospheric turbulence induced distortions \cite{TurbulenceEffects,sunOFDM,CodingFSOMIMO}. Atmospheric turbulence is  caused by pressure and temperature fluctuations in the atmosphere which results in a random behavior in the atmospheric refractive index. Due to AT, the power of an initially transmitted OAM mode leaks to other  modes. This phenomenon causes signal overlapping as well as power disparities between OAM modes.
\newline
To emulate atmospheric turbulence, random phase screens can be placed along the FSO channel. These phase screens are generated based on the modified version of the Kolmogorov spectrum given by \cite{sunOFDM}:
\begin{eqnarray}
\Phi (\kappa )=0.033C^2_n\frac{\exp(-\kappa ^2/\kappa_l^2)}{(\kappa ^2+1/L_0)^{11/6}}f(\kappa,\kappa_l),
\end{eqnarray}
where \mbox{$f(\kappa,\kappa_l)=[1+1.802(\kappa/\kappa_l)-0.254(\kappa/\kappa_l)^{7/6}]$}. $C^2_n$ is the refractive index structure parameter, $L_0$ is the outer scale of the turbulence, $\kappa_l=\frac{3.3}{l_0}$, with $l_0$ is the inner scale of the turbulence. The  turbulence strength in an FSO channel is given by the Rytov variance defined as \mbox{$\sigma _R^2=1.23C^2_n(2\pi/\lambda )^{7/6}z^{11/6}$}, where $C^2_n$ represents the refractive index structure parameter, $\lambda$ is the carrier wavelength and $z$ is the propagation distance. We note that for $\sigma _R^2<1$ ($\sigma _R^2>1$), the system is operating under a weak (strong) turbulence regime. In Fig. 2, we plot, the Phase front of an OAM mode of topological charge $m=+1$ and the corresponding mode purity graphs for different atmospheric turbulence regimes after 1 km of propagation. In the absence of AT, the phase front is not distorted and the initially launched power remains on the same mode. For a weak turbulence regime, the phase front of the OAM mode is affected but can still be recognizable and the spread of the optical power to other modes is low. However, in the strong AT regime, we clearly notice that the phase front of the electromagnetic field is severely impacted. This results in a high power leakage to other OAM modes. In Fig. 3, we plot the cumulative distribution function (CDF) of the average capacity of the transmission of OAM mode $m=+1$ for different turbulence regimes and  SNR= 15 dB. From the figure, we notice, the impact of AT resulting in a reduced channel capacity. 
\begin{figure}[h]
	\centering
	\includegraphics[width=\columnwidth,height=5cm]{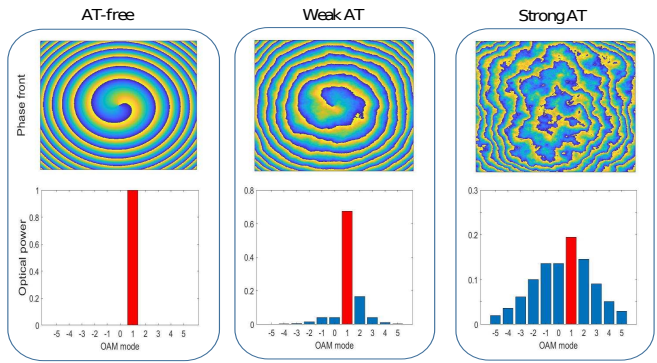}
	\caption{Phase distrortion for OAM mode l=+1 and received optical power for OAM modes for different atmospheric turbulence regimes.}
	\label{Turbulence_states}
\end{figure}
\begin{figure}[h]
	\centering
	\includegraphics[width=\columnwidth,height=5.5cm]{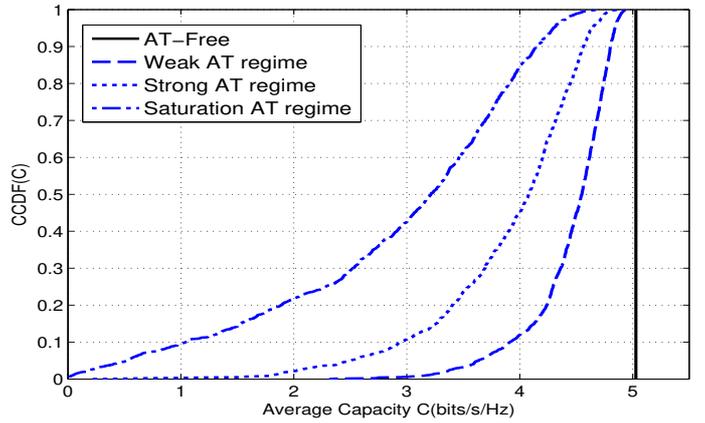}
	\caption{CDF of the average capacity of OAM mode $m=+1$ for different levels of atmospheric turbulence.}
	\label{CDF_AverageCap}
\end{figure}
\section{OFDM-IM Channel Model for OAM FSO Systems}
\subsection{OFDM-IM Coding}

The block diagram of the OFDM-IM scheme is given
in Fig. 1. For the transmission of each OFDM block, a
total of $m$ information bits enter the transmitter of the
OFDM-IM scheme. These bits are split into $G$ groups
each containing $b = b_1 + b_2$ bits, which are used to form
OFDM subblocks of length $N_G = N_{\text{FFT}}/G$, where $N_{\text{FFT}}$ is the size of the fast Fourier transform (FFT). For each sub-block $g\in \left \{ 1, 2, \ldots,G \right \}$, only $K$ out of $N_G$ available subcarriers are activated by the index selector according to the corresponding  $b_1=\left \lfloor \log_2\binom{N_G}{K} \right \rfloor$ bits (see Fig. \ref{channel}). This subcarrier index selection procedure can be performed either using a look-up table for small $N$ and $K$ values or using a one to one mapper based on combinatorial method, which maps natural numbers to $K$-combinations \cite{frenger1999parallel}.
For each sub-block $g$, the remaining $b_2= K\log_2\left ( M \right ) $ bits of the $b$-bits input bit sequence are mapped onto the $M$-QAM signal constellation in order to determine the data symbols that are transmitted over the active subcarriers. 
For example, let us consider $\left ( K,N_G,M \right )=\left ( 1,4,4 \right )$, where only one subcarrier among the four available per the sub-group $N_G$ is modulated by a 4-QAM modulation.  Hence the total transmitted bits per sub-block is equal to four. The first two bits gives fours different possibilities of the index of the modulated subcarrier as shown in Table I. the remaining two bits are used to create the 4-QAM symbols. Hence, the transmitted modulated signal per sublock has 16 different possibilities as: 
\begin{equation}
x_g\in \left \{ \begin{pmatrix}
\pm 1 \pm i\\ 
0\\ 
0\\ 
0
\end{pmatrix},\begin{pmatrix}
0\\ 
\pm 1 \pm i\\ 
0\\ 
0
\end{pmatrix},\begin{pmatrix}
0\\ 
0\\ 
\pm 1 \pm i\\ 
0
\end{pmatrix},\begin{pmatrix}
0\\ 
0\\ 
0\\ 
\pm 1 \pm i
\end{pmatrix} \right \}
\end{equation}
\begin{table}[t]
	\caption{IM table with $K=1$, $N_G=4$, and $b_1=2$ }
	\centering
	\begin{tabular}{|c|c|}
		\hline
		{Incoming bit $b_1$}  &{Active carrier number}  \\
		\hline \hline
		$00$ & $1$ \\
		\hline
		$01$ &  $2$ \\
		\hline
		$11$ &  $3$   \\
		\hline
		$10$ &  $4$  \\
		\hline
	\end{tabular}\label{table1}
\end{table}
\par
After the $M$-QAM modulation is done for all the $G$ blocks, an $N_{\text{FFT}}$ size inverse fast Fourier transform (IFFT) is performed to create the OFDM symbol, and then a cyclic prefix is added to obtain the transmitted vector $\mathbf{x}$ (see Fig. 1). After the signal vector  $\mathbf{x}$ is created, it is used as input data to the electro-optical modulator (EOM) to create an optical signal carried by a Gaussian beam. Afterwards, the latter signal is sent through an SLM to create the desired OAM beam. All OAM signals are then multiplexed and sent through the turbulent FSO channel. At the receiver side, OAM signals are first de-multiplexed and converted back to a Gaussian shape using SLMs, then detected using optical receivers. The offline signal processing is given in Fig. \ref{channel} by the OFDM-IM receiver, where a cyclic prefix removal is followed by an FFT operation and the a maximum likelihood decoding. The equivalent input-output relationship of the transmission chain can be written as:
\begin{equation}
\mathbf{y}=\textbf{h}\odot \mathbf{x}+\mathbf{n},
\end{equation}
where $\odot $ is the element-wise product. $\mathbf{y}$, and $\mathbf{n}$ are the received signal and the noise vector, respectively. The FSO channel $\mathbf{h}$ is computed using the commonly known Split-step Fourier method \cite{sunOFDM}.
After propagation through atmospheric turbulence, the received mode $u_{p,z}$ is detected using the analyzing field of the conjugate of the same OAM mode.
$\mathbf{n}$ is an Additive White Gaussian noise with zero mean and a variance $N_0$. The signal-to-noise ratio (SNR) is defined as $\rho=E_b/N_0$ with $E_b=(N_{\text{FFT}}+N_{\text{CP}})/b$.
\vspace{-15pt}
\subsection{Spectral efficiency}
Spectral efficiency is a relevant performance metric in communication systems. It relates to the bit-rate that can be achieved within a certain
allocated bandwidth $B$.
In our system, $G$ groups of $b =\left \lfloor \log_2\binom{N_G}{K} \right \rfloor + K\log_2\left ( M \right )$ bits per each group  are transmitted during $\Delta T=T_{\text{FFT}}+T_{CP}$. Hence the SE  is given by:
\begin{eqnarray}
\eta=\frac{G\left ( \left \lfloor \log_2\binom{N_G}{K} \right \rfloor + K\log_2\left ( M \right ) \right )}{\left ( T_{\text{FFT}}+T_{\text{CP}} \right )B}.
\label{SE}
\end{eqnarray}  
To maximize the SE, first we fix the number of $G$ groups and derive the optimal value of $K$ that maximizes the bit-rate $b$. 

By ignoring the floor function and by taking the derivative of $b$ with respect to $K$ we obtain: \begin{eqnarray}
\frac{db}{dK} = \log_2\left ( H_{N_G-K}-H_{K} \right )  +  \log_2\left ( M \right ),
\end{eqnarray} 
where $H_{K} = \sum_{i=1}^{K}\frac{1}{i}$ is the harmonic number of order $K$ which can be approximated by $\gamma + \log(K)$, where  $\gamma$ is the Euler constant. By substituting this in the previous equation and making the derivative equal to zero we obtain the optimal number of active subcarriers approximation:
\begin{eqnarray}
K_{opt}\approx \left \lfloor \frac{MN_{\text{FFT}}}{G(M+1)} \right \rfloor.
\end{eqnarray}
\begin{figure}[h]
	\centering
	\includegraphics[width=\columnwidth,height=5.5cm]{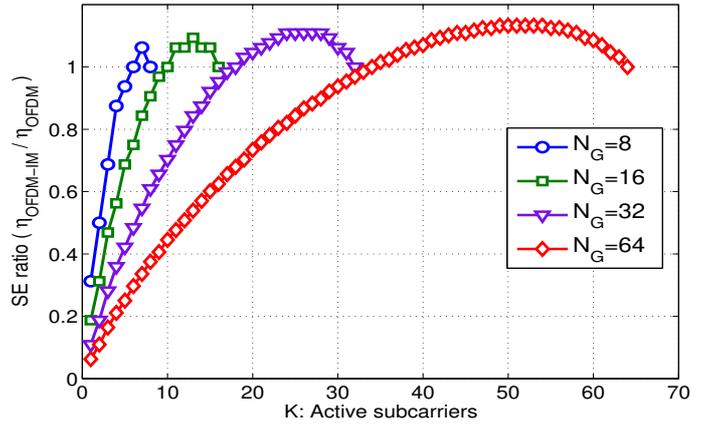}
	\caption{Spectral efficiency ratio of the OFDM-IM scheme to the conventional OFDM for different number of groups $G$}
	\label{SE_groups}
\end{figure}
To have an insight on the  SE improvement of the OFDM-IM scheme, we compute the SE ratio of the OFDM-IM to the conventional OFDM.   
In Fig.~\ref{SE_groups}, we plot the SE  ratio as a function of the number of active subcarriers $K$ per group for $M=4$, $N_{\text{FFT}}=128$, and different number of groups $G$. From the figure, we notice an important SE gain brought by the IM. This gain also increases with an increasing number of total carriers per group.  
\section{BER Performance Evaluation of OAM-FSO System with OFDM-IM}
We assume that we have a perfect knowledge of the channel state information at the receiver, and we consider a maximum likelihood detection (MLD). First, the received OFDM block is split into $G$ groups in order to jointly determine the transmitted active subcarriers and the corresponding modulated symbols. Hence, the received signal for the group $g$ can be written as:
\begin{equation}
\mathbf{y_g}=\textbf{h}_g\odot\mathbf{x}_g+\mathbf{n}_g,
\end{equation}
The MLD of the OFDM-IM scheme makes a joint decision on the active
indices and data symbols for each group $g$ by minimizing the following Euclidean distance: 
\begin{equation}
\hat{\mathbf{x}}_{\text{ML}}=\underset{\mathbf{x} \in \mathfrak{C}}{\text{argmin}}\left \| \mathbf{y}_g-\mathbf{h}_{g}\odot\mathbf{x} \right \|^2,
\end{equation}
where $\mathfrak{C}$ is the set of all possible combinations of modulated active subcarriers, with $\mathfrak{C}=2^{b_1}\times M^K$. The term $2^{b_1}$ refers to the total number of possible active subcarriers combinations and $M^K$ is the total possible combinations for the modulated $K$ subcarriers.
In the following, we suppose that $\mathbf{h}_g$ is known at the receiver.  Let  $\mathbf{x}_i$ ($\mathbf{x}_j$) be the transmitted  (the estimated) codeword. The error probability is defined as:
\begin{equation}
	P_{\text{e}}=\sum_{\mathbf{x}_i\in \mathfrak{C}}{} \text{P}_\text{r} \left \{ \mathbf{x}_i \right \}\text{P}_\text{r} \left \{ \mathbf{x}_j\neq  \mathbf{x}_i  \mid \mathbf{x}_i \right \}
\end{equation}
By using the union bound of the error probability \cite{javed2019asymmetric}, we obtain:
\begin{equation}
	P_{\text{e}} \leq \frac{1}{\text{card}(\mathfrak{C})} \sum_{\mathbf{x}_i \neq  \mathbf{x}_j} \text{P}_\text{r} (\mathbf{x}_i, \mathbf{x}_j),
\end{equation}
where $\ \text{P}_\text{r} (\mathbf{x}_i, \mathbf{x}_j)$ is the pairwise error probability (PEP) of  transmitting $\mathbf{x}_i$ and decoding $\mathbf{x}_j$. The evaluation of the error probability simplifies to the averaging of the conditional PEP given by the $Q$-function as: 
\begin{equation}
	\text{P}_\text{r} (\mathbf{x}_i, \mathbf{x}_j/\mathbf{h}_g)=Q\left( \sqrt{\frac{\left \| \mathbf{h}_g \odot \left ( \mathbf{x}_i-\mathbf{x}_j \right ) \right \|^2}{2N_0 }}\right), 
	\label{eq.pep}
\end{equation}
The derivation of a closed form expression for Eq. \eqref{eq.pep} requires the knowledge of the statistics of the OAM-FSO channel $\mathbf{h}_g$ which are not known at the moment. Thus, we  evaluate the BER performance through Monte Carlo simulations.   
\subsection{BER Performance Evaluation }
In this section, we evaluate the BER performance of the OFDM-IM and OFDM schemes. We simulate the propagation of OAM beams for the weak and strong atmospheric turbulence regimes. The beam waist at the transmitter for all beams is set to $\omega_0=1.6$ cm to ensure a minimal beam waist at the receiver plane. On the other hand, we assume that the optical receiver is large enough to collect all received OAM beams. To create the desired OAM modes, SLMs with LCD of dimension \mbox{$512 \times 512$} pixels are used.  The propagation distance is set to $z=1~$km, the inner and outer scales of turbulence are set to $l_0=5~$mm and $L_0=20~$m, respectively. AT is emulated by placing $20$ random phase screens each  $50~$m. Each phase screen is evaluated  as the Fourier transform of a complex random distribution with zero mean and variance equal to  $\left ( \frac{2\pi}{N\Delta x} \right )^2\Phi (\kappa )$, where $N=512$ is the array length and $\Delta x=5~$mm is the grid spacing assumed to be equal in both dimensions $x$ and $y$. Propagation through the turbulent atmosphere is simulated using the commonly used split-step Fourier method at wavelength \mbox{$\lambda=1550$ nm}. For weak (strong) turbulence, we set $C^2_n=10^{-14}$ ($C^2_n=10^{-13}$), respectively. The transmitted signals are either OFDM or OFDM-IM having the following parameters: $N_{\text{FFT}}=128, M=4, N_{G}=4, K=1$.
In Fig.~5 and Fig.~6, we compare the BER performance of the OFDM and OFDM-IM for the weak and strong turbulence regimes by considering OAM modes $m\in \left \{ +1, +3 \right \}$. From the figures, we notice that for both turbulence regimes a considerable SNR gain is obtained for the OFDM-IM compared to the classical OFDM scheme.
\begin{figure}[h]
	\centering
	\includegraphics[width=\columnwidth,height=5.5cm]{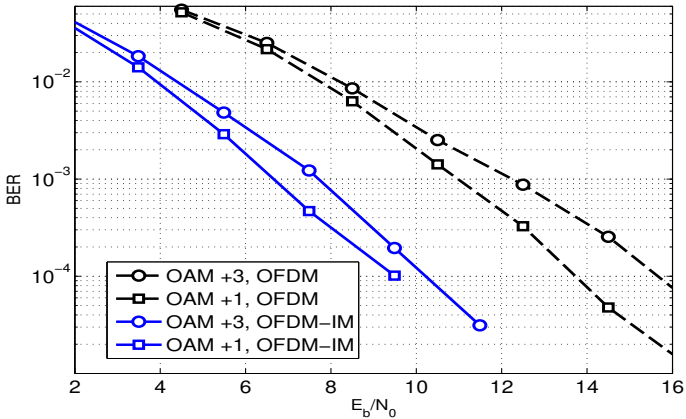}
	\caption{BER vs SNR comparison for OFDM and OFDM-IM and OAM modes $m\in \left \{ +1, +3 \right \}$ in the weak turbulence regime.}
	\label{BER weak regime}
\end{figure}
\begin{figure}[h]
	\centering
	\includegraphics[width=\columnwidth,height=5.5cm]{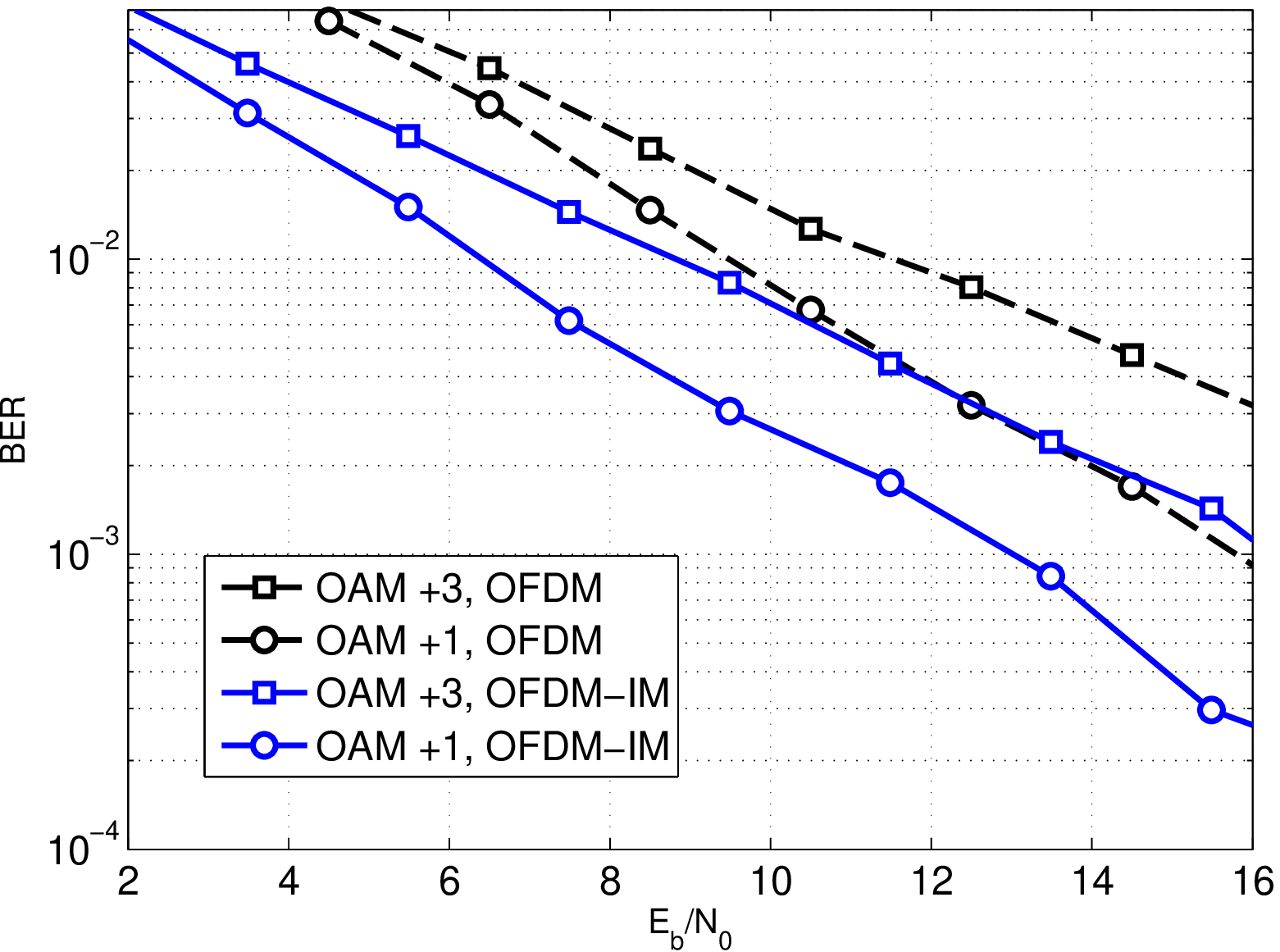}
	\caption{BER vs SNR comaprison for OFDM and OFDM-IM and OAM modes $m\in \left \{ +1, +3 \right \}$ in the strong turbulence regime.}
	\label{BER strong regime}
\end{figure}
\section{Conclusion}
In this paper, we have introduced OFDM with index modulation for OAM transmission in FSO links suffering from atmospheric turbulence. The OFDM-IM scheme was shown to bring important spectral efficiency improvement. Furthermore, OFDM-IM shows significant BER improvement  compared to the classical OFDM format. This significant performance enhancement results encourage further studies on the application of more sophisticated OFDM-IM formats to OAM-FSO systems.

\bibliographystyle{ieeebib}
\bibliography{GL2020}

\end{document}